\documentclass[12pt]{article} 
\input epsf.tex 
\def\be{\begin{equation}} \def\ee{\end{equation}}
\def\bi{\begin{itemize}} \def\ei{\end{itemize}}
\def\bea{\begin{eqnarray}} \def\eea{\end{eqnarray}} \def\ba{\begin{array}}
\def\ea{\end{array}} \def\ben{\begin{enumerate}} \def\een{\end{enumerate}}

\newcommand{\eqn}[1]{(\ref{#1})}

\newcommand{\prl}[3]{Phys. Rev. Lett. {\bf#1} ({#2}) {#3}}

\newcommand{\hepth}[1]{{\tt arXiv:{#1}[hep-th]}}
\newcommand{\arxiv}[1]{{\tt arXiv:{#1}[hep-th]}}

\def\br{\nonumber\\}

\begin{document}
{}~
\hfill \vbox{
\hbox{arXiv:2111.nnnn} 
\hbox{10/11/2021}}
\break

\vskip 3.5cm
\centerline{\large \bf
The information horizon entropy}
\centerline{\large \bf
for quantum dot and a symmetrical bath}

\vskip 1cm

\vspace*{1cm}

\centerline{\sc  Harvendra Singh }

\vspace*{.5cm}
\centerline{ \it  Theory Division, Saha Institute of Nuclear Physics (HBNI)} 
\centerline{ \it  1/AF Bidhannagar, Kolkata 700064, India}
\vspace*{.25cm}

\vspace*{.5cm}

\vskip.5cm
\centerline{E-mail: h.singh [AT] saha.ac.in }

\vskip1cm
DRAFT : \today \\

\centerline{\bf Abstract} \bigskip

We study the entropy of quantum-dot system in contact with 
a symmetrical CFT bath living on the boundary of pure
AdS3 black hole. The q-dot is localized at the centre of 
bath system of finite size. We first determine the exact location of the
`information horizon' for q-dot and then obtain corresponding generalised 
entropy of q-dot plus bath system.
It is done by finding codim-2 time extremal curve whose end point 
uniquely determines the information horizon  of localised
q-dot system. By including the (bulk) entropy contribution
of the information horizon the Page curve for the radiation follows. 
These results can be easily generalized to higher dimensional cases
as well.  
\vfill 
\eject

\baselineskip=16.2pt


\section{Introduction}

 The holographic principle in string theory  \cite{malda} has 
produced easy to understand answers for some of the difficult
and intractable questions in strongly coupled quantum field theories. 
We focus on the phenomenon of information exchange between 
two identical quantum systems having a common interface. 
The  information sharing  is 
a real time phenomenon as quantum states would always get entangled. 
In quantum mechanical theories the information  contained in a given state
cannot be destroyed, cloned or mutated. For example, in the
bi-partite systems the
information can either be found in one part of the Hilbert space 
or in its compliment; see \cite{pati,pati1}.
Generally it is believed that the exchange and sharing of 
quantum information is guided by the unitarity and locality. 
The time like flows of isolated quantum systems should essentially be
Hamiltonian flows. Under similar claims for the black holes the
formation and evoparation processes (via Hawking radiation) it is generally
expected that the enropy curve for the
radiation should bend after half Page-time is crossed \cite{page}. 
This  certainly  holds good when a pure state is divided into two smaller systems. 
But for a mixed state or finite temperature CFT duals to AdS-black holes,
it is not  straight forward to answer this question. 
However, in some recent models by coupling holographic
CFTs to an external radiation (bath) system, and also by
involving  nonperturbative  techniques such as 
replica, wormholes and islands \cite{almheri,replica19}, some
answers to these difficult questions have been attempted.
\footnote{ See a review on information paradox along
different paradigms in \cite{raju}
and for list of major references therein, also in [\cite{susski}-\cite{hashi}].} 

The recent proposal for generalised entanglement
entropy \cite{almheri} involves an hypothesis of the island $(I)$ contribution, 
including the contribution from the island boundary $(\partial I)$, 
such that the complete quantum entropy of radiation (bath) 
can be expressed as
$$ S_{Rad}^A=_{min}[ ext\{ {Area(\partial I)\over 4G} + S[A ~U ~I]\}]  .$$
It means one needs to pick the lowest contribution out of a set of extremas. 
This complicated looking 
formula seemingly reproduces a Page curve for the radiation entropy. 
However one of the puzzling feature of above proposal entails in the ad hoc 
appearance of an island in the black hole geometry usually outside of the horizon. 
The island does not arise by means of a sound dynamical principle. It is merely
presumed to be there, perhaps associated with the presence of the bath system. 
In contrast, in the present work we would
like to propose an alternative picture that there will always exist 
an `{\bf information horizon}' for a quantum-dot living 
at the interface of a symmetrical CFT (bath) system. 
The information horizon is always found to be outside of the black hole horizon. 
Next the information horizon can be dynamically obtained by extremizing 
a {\it time}-curve corresponding to the quantum dot situated 
at the bath interface. Further,
the net information processed by the quantum dot 
in a given time interval remains a well defined physical observable.
\footnote{In a recent work \cite{hs2020}  a measure for net exchange of
 quantum information between two {\it adjacent} subsystems 
(having  common interface) for the 
 dual field theories of $AdS_{d+1}$ was discussed. 
As the quantum systems continuously exchange information  a large
 amount of information will be exchanged over long periods. 
This leads to overall information growth in time. The information
exchange typically grows as 
$\propto({1 \over \delta^{d-2}}- {1 \over t^{d-2}})$ 
for  extremal  $CFT_d$. However over finite time interval only
optimum information is exchanged. 
This  leads to the formation of  `information horizon'. }
Effectively the net information processed is
a measure  of unitary operations (optimal count) a quantum computer 
(like q-dot) would perform in a given time. 
This can be evalauted holographically 
by embedding a codimension-2 time curve
in  asymptotically AdS spacetime,  as described in \cite{hs2020}. 
(The proposal is very much similar to the holographic
 measure of entanglement entropy \cite{RT,HRT}.)
The end point (or cusp-point) 
of time  extremal curve is always unique and it will be identified as the 
information horizon $(I_h)$.   
Correspodingly we propose that there would be a bulk contribution to the entropy
arising out of information horizon of a quantum dot (being in contact with bath).
So that total entropy is given by $S[I_h]+S[A]$. 
Therefore for the quantum entropy of radiation the formula
may be written as

$$ S_{Rad}^A=_{min}\{{Area[I_h]\over 4G} , S[A]\}   $$

The above expression reproduces the Page curve for the radiation. 
We indeed find that when a finite temperature bath becomes sufficiently   
large, it requires q-dot very long time to process entire information 
and correspondingly the cusp-point tends to merge 
with the black hole horizon, i.e. $z_i\to z_0$ in the large bath limit. 
So that the entropy of radiation  becomes 
\bea
S_{Rad}\to S_{BH}
\eea
after a very long time period.

The paper is organized as follows. In section-2 we introduce the information horizon idea and obtain the generalised entropy
formulation for pure $AdS_3$ case. On the boundary we have taken a quatum-dot in contact with finite size symmetrical radiation bath.
We then extend our results for the BTZ black holes case in section-3. The last section-4 contains a brief summary of our observations.   

\section{Information processing by a quantum dot}
An immediate goal in this section is to know how much 
information a quantum dot can process in given time interval. 
We also assume that q-dot is in contact with a symmetrical radiation 
bath of finite size. 

Let us consider  pure $AdS_3$ spacetime geometry
\bea\label{ads3}
ds^2={L^2 \over z^2} (- dt^2 + dx^2 + dz^2)
\eea
where $L$ is the radius of curvature. The coordinate range  $0\le z \le \infty$ 
represents the full holographic range. The Kaluza-Klein
compactification on a circle ($x\simeq x+ 2\pi R$)
 produces a
{\it near} $AdS_2$  solution,  well known as Jackiw-Teitelboim 
 background \cite{JT,JT1},
\bea\label{gy67}
&& ds^2_{JT}={L^2 \over z^2} (- dt^2 + dz^2)\br
&& e^{-2 (\phi-\phi_0)}= \sqrt{g_{xx}}={L \over z} 
\eea
where $\phi$ is the $2d$ dilaton field, written in standard convention 
(effective string coupling vanishes near the boundary). The Newton's constants
get related as ${2\pi R \over G_3}\equiv {1 \over G_2}$, 
with $G_2$ being dimensionless.
 The anti-de Sitter solution without the dilaton 
 remains a topological spacetime with no propagating degrees of freedom. 

\begin{figure}[h]
\centerline{\epsfxsize=3in
\epsffile{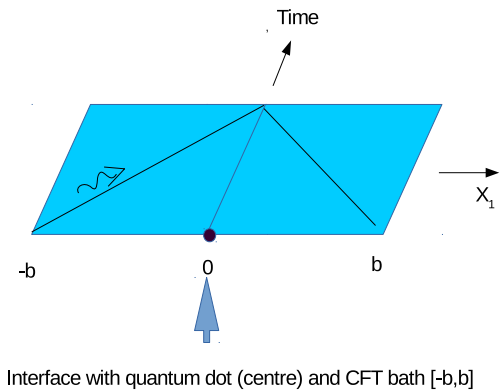} }
\caption{\label{fig21b} 
\it A typical arrangement of quantum dot (at the centre of $x$-axis) and 
1-dimensional radiation bath $x=[-b,b]$. The boundary theory has usual 
Minkowski spacetime. The light signal from the edge of the
bath takes time $b$ to reach at the centre (quatum dot). 
This is the maximum time window relevant for processing of bath
information by a q-dot.}
\end{figure} 
\begin{figure}[h]
\centerline{\epsfxsize=3in
\epsffile{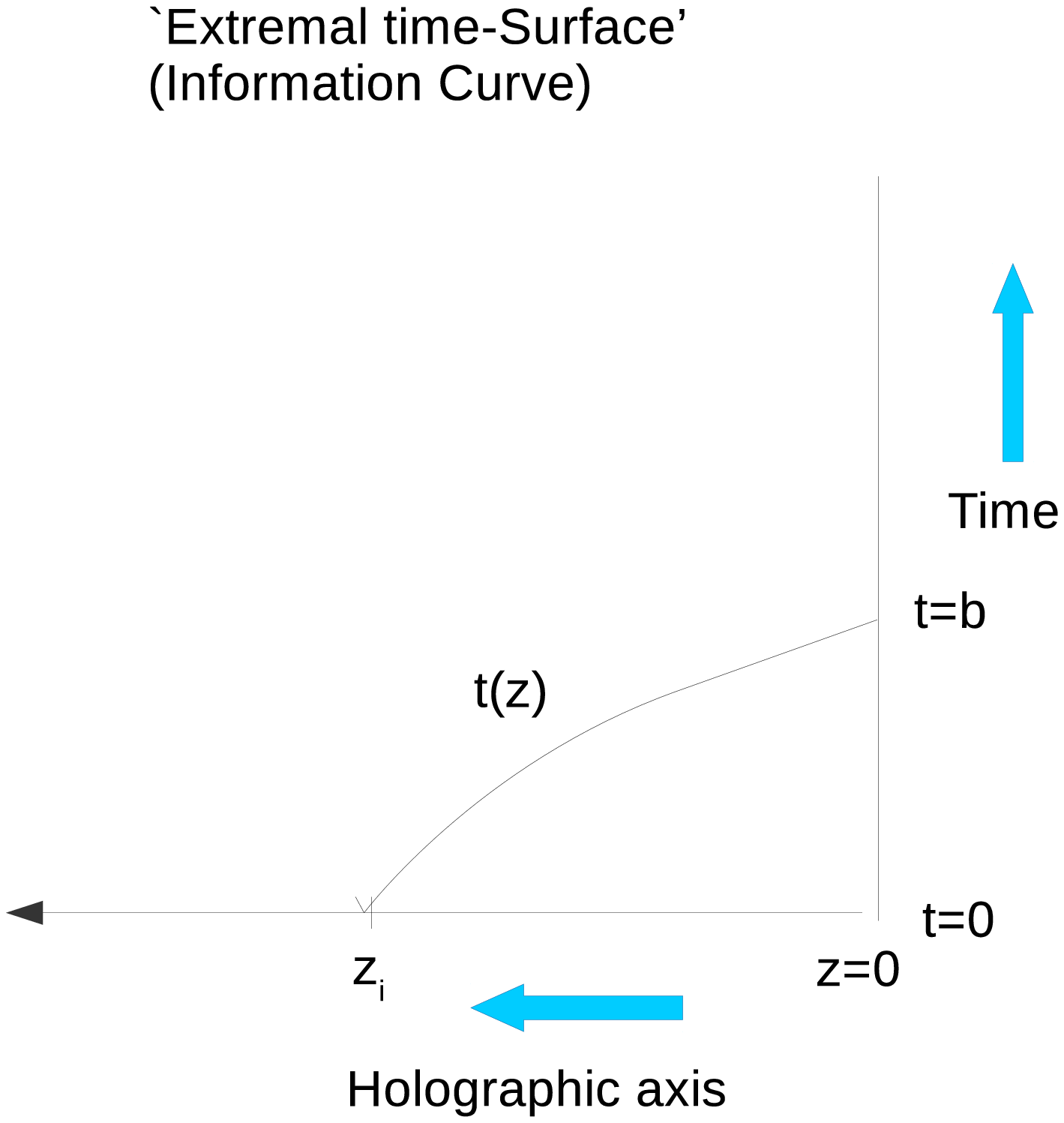} }
\caption{\label{fig21a} 
\it The extremal surface representing the time embedding
in the  $AdS_3$ bulk for specific time interval. The cusp point $z_i$ marks
the end point of time extremal curve. It precisely is the information horizon
of a quantum dot situated at the bath interface. }
\end{figure} 

The $CFT_2$ lives on entire 2-dimensional noncompact 
$(t,x)$ flat  boundary of $AdS_3$ geometry.
We consider a CFT subsystem  $[-b,b]$ along spatial $x$ direction with
an interface at the centre $x=0$. At the interface 
we place a defect system (a quantum dot)  
that is described by a quantum mechanical
theory of its own, see figure \eqn{fig21b}. To be consistent with the
bath $CFT_2$
 the quantum theory is necessarily taken to be  "near $CFT_1$", 
  a holographic  dual of the JT-gravity as described in \eqn{gy67}. 
Due to this we safely assume  that any back reaction of  quantum dot  
on bath CFT is ignorable and vice versa. 
The entire systems set up is taken in a particular symmetrical way
for the conveniece. The states of the q-dot and the  bath 
are necessarily entangled. 

It is known that the entanglement entropy of extremal $CFT_2$  of size $2b$ 
is ordinarily given by 
\bea
S_{bath}= {L \over 2 G_3}\ln{2b \over \epsilon}
\eea
Since there is also a point like quantum
system at the centre and its states are entangled with the bath,
the information will be processed by the quantum dot based on its
independent dynamics. We are assuming a unitary set up here. 
The net window of time we are interested 
in is however fixed by the maximum time the radiation takes to travel
from one edge of the bath subsystem to the q-dot located at its centre $x=0$. 
Simply the maximum time window relevant for q-dot (unitary) operations is  
\bea
\bigtriangleup t_{max}=b
\eea 
Note the maximum information that can be processed 
by a quantum dot in given time would always be finite \cite{hs2020}.
Holographically, as proposed in \cite{hs2020}, 
the entanglement information processed (through combinations of multiple unitary operations) 
by any quantum system may be obtained  by extremizing an
action functional  as
\bea\label{ki9}
I_E={L\over 2 G_3}\int_\epsilon^{z_i} {dz\over z}\sqrt{1-(\partial_z t)^2} 
\eea
where $z=z_i$ is the location of the cusp and $\epsilon$ is the UV cut-off. 
The integral expression describes the area of codimension-2 curve, for constant $x$
(with $x=0$), in
$AdS_3$ geometry \eqn{ads3}. The cusp is an end point of 
the extremal curve  inside the bulk geometry, shown in the figure \eqn{fig21a}. 
The equation of an extremal time curve  from \eqn{ki9} is given by
\bea 
{t'}= {z\over z_i \sqrt{1+{z^2\over z_i^2}} }
\eea
where b.c.s are:  $t'|_{z=0}=0$ and at the cusp point $t'|_{z=z_i}={1\over \sqrt{2}}$.
By integrating this equation we obtain an exact answer
\bea
\bigtriangleup t=   B_0 z_i
\eea
 where constant  $(B_0=\sqrt{2}-1\simeq .414$). 
Thus the location of the cusp point for the bath parameter $b$  is precisely 
\bea\label{isl1}
z_i={\bigtriangleup t_{max}\over B_0}={b\over \sqrt{2}-1}
\eea
where $\bigtriangleup t_{max}$ is the time required by a signal to reach
at the location of the quantum dot at the centre of bath from its boundary.
Note that from eq.\eqn{isl1} we get $z_i>b$ under all situations. 
The cusp point $z_i$ may also be taken as an end point of information wedge  $[0,z_i]$  with other end at the AdS boundary.
In th next, we now claim that $z_i$ is the information horizon $(I_h)$ 
of the q-dot system. If the bath size $2b$ increases correspondingly 
the location of the information horizon will also change in tandem. 
The bulk region $z_0 \ge z >z_i$, inside information horizon, may then describe 
an {\bf island} for complementary CFT system ($[b,\infty]$ plus $[-b,-\infty]$). 
Although there is practically no need of any physical
islands in our bath plus q-dot system set up. 
{\it Ultimately the extremality of the information  quantity ($I_E$)  
determines the location of information horizon.} 
This proposal is distinctly different to the similar 
set up \cite{almheri} where maximality of a generalised entropy 
fixes the location of an island boundary.
Although surprisingly we find that in both the procedures one gets $z_i>b$. 

The gravitational entropy corresponding to information horizon located at $z=z_i$
can be written as
\bea\label{plk1}
S_{bulk} [I_h]&=& {l_x\over 2 G_3}{ L \over z_i} + constant = {l_x L B_0\over 2 G_3 b} + constant 
\eea
Here we take $l_x \gg b$ as it describes the IR scale of $CFT_2$. 
The  $[-l_x,l_x]$ covers the entire range of CFT $x$ coordinate. (If $x\sim x+2\pi R$   
is compactified then we must take $l_x = \pi R$.) 
 Thus the entropy contribution due to quantum dot  corresponding to its
information horizon is given by
\bea 
S_{dot} \equiv S[I_h].
\eea
A generalised entropy of the bath  and  quantum dot together may be written as 
\bea
S_{gen}&&= S_{dot} + S_{bath}\br
&&={ L\over 2G_3}\left({ l_x B_0\over b} +  \ln {2b\over\epsilon}\right)+ constt.
\eea
where $\epsilon$ is UV cut-off term for both
q-dot system and the bath. It is rather useful to define a dimensionless variable 
$\tilde b=b/l_x$,
 keeping in mind the hierarchy of scales $l_x \gg b > \epsilon$,
we  reexpress
\bea
S_{gen}={ L\over 2G_3}\left({  B_0\over \tilde b} +  \ln {2\tilde b} \right)+ S_0.
\eea
The overall constant  $S_0\approx O( \ln {l_x\over \epsilon}) $ 
and contains other parameters.
Obviously the generalised entropy varies with bath parameter $b$ and there 
is a unique minimum for the total entropy 
at $\tilde b=B_0$.
However, our claim is that
 the  quantum entropy of the radiation would be only given by   
(for any $b$) by the following selection rule
\bea\label{fin1}
S_{Rad}=_{min}\{S_{I_h}, S_{bath}\}
\eea 
This expression makes the quantum radiation entropy proposal complete and no
further extremization is necessary. Interestingly, the 
gravitational entropy of information horizon (relevant for quantum-dot)
always dominates for small bath size,
whereas the bath term dominates when $b$ becomes sufficiently large.
For larger bath sizes the information horizon ($I_h$) contribution 
becomes subleading. A plot
has been drawn for various entropy components in figure \eqn{fig5aa}
There is a crossover point where
two contributions become exactly equal.  
Hence there is a Page curve phenomenon here for $S_{Rad}$
provided we keep the contribution only of the smaller component
in $\{S_{dot}, S_{bath}\}$.
 The Page curve \cite{page} for radiation
follows from the standard
principle that the minimum entropy is to be favoured.
 \begin{figure}[h]
\centerline{\epsfxsize=3.5in
\epsffile{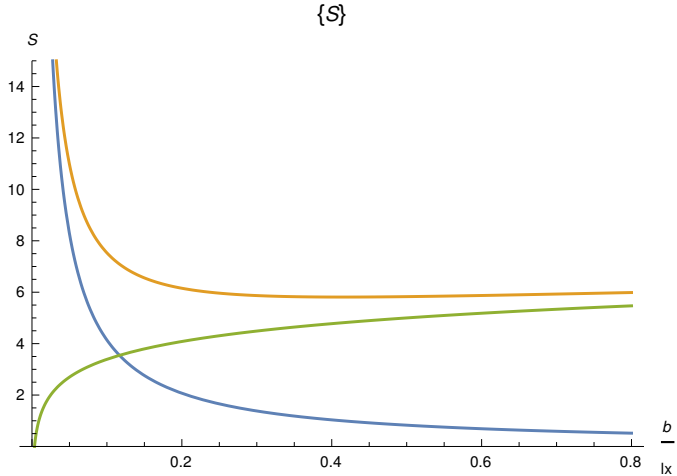} }
\caption{\label{fig5aa} 
\it The blue curve is for information horizon entropy
for quantum dot. The rising graph (green) is for the radiation entropy.
The  topmost graph represent total of two entropies. 
We set $l_x=1, {L\over 2 G_3}=1$. There is a minimum at $\tilde b=B_0\sim.414$ for
the total entropy.}
\end{figure}
 That is so far for extremal AdS case of  zero temperature 
CFT  with a quantum dot in a symmetrical bath set up.

{\it Some Comments:}
Our result  seems to provide  an alternative way to understand  radiation 
bath models developed recently, see ref.\cite{almheri},
where the  presence of a bulk island $(I) $ was expected but the location of
island boundary ($\partial I$) was largly kept arbitrary. 
Although there appears to be a lack of clear 
holographic understanding in determining the bulk island. 
As an alternative we have obtained the information horizon ($I_h$) holographically
by extremizing codim-2 time-curve for the q-dot information processing.
The island models rely primerily on the assumption that island will be there,
but anywhere
in front of the horizon (even if sometime behind the horizon). 
Furthermore this requires  to extremize
a generalised entropy (bulk entropy together with q-dot plus bath entropy) 
all by hand (not following holographic or dynamical principle). 
In contrast we have proposed here that the `information 
horizon' holographically appears when we extremize the information processed
by the quantum dot. The
information time-curve entirely encodes the information processing capacity
(through unitary operations)
of a quantum mechanical device (described by dual JT gravity) and in contact with finite size bath. 

\subsection{An equilibrium in time?}
Since the expression \eqn{fin1} is true for any size ($b$) of the bath CFT, 
and that $\bigtriangleup t_{max}=b$, we may convert above result \eqn{plk1}
into a time dependent expression, namely
\bea
S_{gen}= {l_x L B_0\over 2 G_3 \bigtriangleup t_{max}} + { L \over 2 G_3}
\ln ({2\bigtriangleup t_{max}\over\epsilon})
\eea
Using this a Page time graph can be deduced ($l_x> \bigtriangleup t>\epsilon$). 
For small time interval the gravitational contribution 
(q-dot) leads  whereas the entropy of the radiation 
(bath) remains smaller, but the radiation contribution
 starts rising for large time gap and keeps on growing 
until $T_{Page}$. After the Page time radiation entropy crosses 
the geometric contribution of the information horizon. 
The crossover point may not generally be the point where 
the total  entropy will be the lowest. The crossover may actually be described as 
the point of equilibrium where information processing 
by the dot system is in equilibrium with the bath system.

\section{Radiation entropy at finite temperature}
   For finite temperature entropy the spacetime AdS geometry will be 
taken with a horizon
\bea\label{btz23}
ds^2={L^2 \over z^2}
(-f dt^2 +{dz^2 \over f } + dx^2)
\eea
The function $f(z)=(1-{z^2\over z_0^2})$ and
 $z=z_0$ is location of black hole horizon.
There is finite temperature in the field theory at boundary.
\footnote{One may set coordinate $x=L\phi$, with range $0\le \phi\le 2\pi$ for BTZ
black hole.}  Now
the quantum dot is taken in thermal equilibrium with symmetrical bath
and is located at the centre $x=0$
of CFT subsystem of size $[-b,b]$. 
 
We embed the time coordiante of the quantum dot system 
inside the bulk geometry \eqn{btz23}.
Correspondingly the information action for q-dot located at $x=0$ is
given by
\bea\label{ki9a}
I_E={L\over 4 G_3}\int_\epsilon^{z_i} {dz\over z}\sqrt{{1\over f}-f(\partial_z t)^2} 
\eea
From this we  get the following equation describing an extremal time curve
\bea\label{fbn}
t'= {z/z_i\over f \sqrt{f+(z/z_i)^2}}
\eea
where $z=z_i$ is the cusp point and it corresponds to  value
$t(z_i)=0$ at the boundary. 
Especially the slope of  extremal curve at the cusp point is
\bea 
{t'}|_{z=z_i}= {1\over  (1-{z_i^2\over z_0^2})
\sqrt{2-{z_i^2\over z_0^2}}} \ge {1\over\sqrt{2}}
\eea
Note that this slope in the black hole case is always larger than  
${1\over\sqrt{2}}$, a value  we have found for extremal $AdS_3$ 
background in the previous section.
The slope vanishes near the boundary $z=0$.
From here we  get an exact expression for
 $z_i$ in terms of interval $\bigtriangleup t$.  By integrating \eqn{fbn},
it is given by 
\bea\label{ki10}
\tanh ({\bigtriangleup t \over z_0})= {{z_i\over z_0}
(\sqrt{2-{z_i^2\over z_0^2}}-1)\over
1-{z_i^2\over z_0^2}\sqrt{2-{z_i^2\over z_0^2}} } 
\eea
and equally well, by knowing $\bigtriangleup t_{max}=b$, it
can be written as
\bea\label{ki11}
\tanh ({ b \over z_0})= {{z_i\over z_0}
(\sqrt{2-{z_i^2\over z_0^2}}-1)\over
1-{z_i^2\over z_0^2}\sqrt{2-{z_i^2\over z_0^2}} } 
\eea
A plot for $\bigtriangleup t ~ Vs ~z_i$ has been provided 
in the figure \eqn{fig211}
for the convenience.

The net information processed by a quantum dot in  given time interval
$\bigtriangleup t_{max}=b$ remains always finite \cite{hs2020}.
Consequently this leads to the existence of an information horizon at $z=z_i$.  
 The gravitational entropy corresponding to the information horizon 
is
\bea
S[I_h]= {L \over 2G_3} {l_x \over z_i } + const.
\eea
and this is the entropy of the quantum dot
\bea
S_{dot}= S[I_h]\ ,
\eea
where $z_i$  is to be determined from eq.\eqn{ki11} for given $b$. 
The range $[-l_x,l_x]$ determines the box size of $x$ coordinate.
(But $l_x= \pi L$ for using it for BTZ case.)
The  entropy of the finite temperature
radiation bath of size  $2b$ is given by 
\bea
S_{bath}= {L\over 4G_3} \ln \sinh^2({2b\over z_0})  
\eea
Hence the generalised entropy at finite temperature for the combined system
will be 
 \bea\label{fin1a}
S_{gen}=  {L \over 2G_3}\left({l_x \over z_i } +   \ln \sinh{2b\over z_0}
 \right).
\eea
While the quantum entropy of the radiation would be the minimum of the two
values
 \bea\label{fin17}
S_{Rad}=_{min} \{S_{dot}, S_{bath}\}
\eea
\begin{figure}[h]
\centerline{\epsfxsize=3in
\epsffile{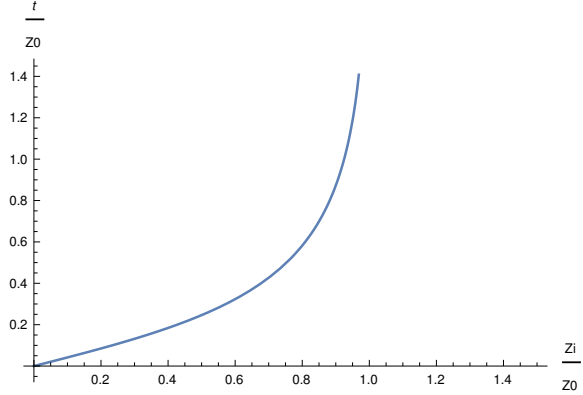} }
\caption{\label{fig211} 
\it The $\bigtriangleup t ~Vs~ z_i$ plot for BTZ black hole. 
When $z_i \to z_0$ the  time required for information 
processing by quantum dot indeed becomes very large.
We have taken $z_0=1$ here.  }
\end{figure} 

For small bath cases, i.e. $b\ll z_0$, from eq.\eqn{ki11} one obtains $ b\simeq z_i (\sqrt{2}-1) + O({z_i^3\over z_0^3})$. This gives  zero temperature 
result once higher orders are neglected and it is expected. In other words the black hole horizon does not play significant role in the entanglement for very small bath subsystems. The quantum entropy of the radiation can still be 
approximated as $\sim {L \over 2G_3}\ln {2b\over z_0}$.
However  under  large bath limit ${b\to \infty}$, equally
for large  time intervals (as $ \bigtriangleup t_{max} \to\infty$), we 
 indeed find from \eqn{ki11} that it leads to $z_i \to z_0$, 
 the information horizon tends to merge with 
the BH horizon. Thus the geometric entropy contribution  of information horizon becomes  
(for ${b\to \infty}$ )
\bea
S_{dot} \to   {\pi L^2 \over 2G_3 z_0}\equiv S_{BH}
\eea
which is the lowest value of the entanglement entropy at finite temperature. 
Thus from \eqn{fin17} for large bath case the quantum entropy of radiation
would become
\bea
Lim_{b\to \infty} S_{Rad} \simeq S_{BH}.
\eea
  
\section{Summary}
We have explored the entanglement dynamics of a quantum mechanical 
system in contact with 
1-dimensional radiation bath. The  bath has an interface at $x=0$
where the q-dot is located. The strongly coupled quantum system is living on the boundary of
$AdS_3$ spacetime, including the black holes. There is an entanglement and
 information exchange (sharing)  between two subsystems 
at the interface. 
We  first calculated the precise location of an information horizon
inside the bulk geometry. 
We do it by extremizing codim-2 time-curve whose end point in the bulk 
uniquely determines the location of the information horizon. 
The information horizon is found to be located always outside  the
BH horizon. We have calculated the genralized entropy of q-dot and the bath.
It is proposed that quantum radiation entropy follows the principle that
$$ 
S_{Rad}=_{min} \{S_{dot}, S_{bath}\}
$$
We also find that under large bath limit the quantum entropy of radiation
tends to become
$$
S_{Rad}\simeq S_{BH}
$$
and thus realising the Page curve for the entropy of thermal radiation.  
These results can be generalized for
higher dimensional CFT cases also.


\vskip.5cm
   


\end{document}